\documentclass[manuscript,screen]{acmart}

\usepackage{balance}

\AtBeginDocument{%
  \providecommand\BibTeX{{%
    \normalfont B\kern-0.5em{\scshape i\kern-0.25em b}\kern-0.8em\TeX}}}

\emergencystretch=\maxdimen
\copyrightyear{2021}
\acmYear{2021}
\setcopyright{rightsretained}
\acmConference[CSCW '21 Companion]{Companion Publication of the 2021 Conference on Computer Supported Cooperative Work and Social Computing}{October 23--27, 2021}{Virtual Event, USA}
\acmBooktitle{Companion Publication of the 2021 Conference on Computer Supported Cooperative Work and Social Computing (CSCW '21 Companion), October 23--27, 2021, Virtual Event, USA}
\acmDOI{10.1145/3462204.3481743}
\acmISBN{978-1-4503-8479-7/21/10}

\begin{document}

\title{Authoring Platform for Mobile Citizen Science Apps with Client-side ML}


\author{Fahim Hasan Khan}
\email{fkhan4@ucsc.edu}
\orcid{0000-0003-3130-6259}
\affiliation{%
  \institution{University of California, Santa Cruz}
%  \streetaddress{ }
  \city{Santa Cruz}
  \state{CA}
  \country{USA}
  \postcode{95064}
}
\author{Akila de Silva}
\email{audesilv@ucsc.edu}
\affiliation{%
  \institution{University of California, Santa Cruz}
%  \streetaddress{ }
  \city{Santa Cruz}
  \state{CA}
  \country{USA}
  \postcode{95064}
}

\author{Gregory Dusek}
\email{gregory.dusek@noaa.gov}
\affiliation{%
  \institution{NOAA National Ocean Service}
%  \streetaddress{ }
  \city{Silver Spring}
  \state{MD}
  \country{USA}
  \postcode{20910}
}

\author{James Davis}
\email{davisje@ucsc.edu}
\affiliation{%
  \institution{University of California, Santa Cruz}
%  \streetaddress{ }
  \city{Santa Cruz}
  \state{CA}
  \country{USA}
  \postcode{95064}
}

\author{Alex Pang}
\email{pang@soe.ucsc.edu}
\affiliation{%
  \institution{University of California, Santa Cruz}
%  \streetaddress{ }
  \city{Santa Cruz}
  \state{CA}
  \country{USA}
  \postcode{95064}
}

\renewcommand{\shortauthors}{F.H. Khan, A. de Sliva, G. Dusek, J. Davis, and A. Pang}

\begin{abstract}
Data collection is an integral part of any citizen science project. Given the wide variety of projects, some level of expertise or, alternatively, some guidance for novice participants can greatly improve the quality of the collected data. A significant portion of citizen science projects depends on visual data, where photos or videos of different subjects are needed. Often these visual data are collected from all over the world, including remote locations. In this article, we introduce an authoring platform for easily creating mobile apps for citizen science projects that are empowered with client-side machine learning (ML) guidance. The apps created with our platform can help participants recognize the correct data and increase the efficiency of the data collection process. We demonstrate the application of our proposed platform with two use cases: a rip current detection app for a planned pilot study and a detection app for biodiversity-related projects.
\end{abstract}





\maketitle

\section{Introduction and Background}

Crowdsourcing is a distributed task assignment model where large groups of paid or unpaid participants submit their works, typically some form of data, using the internet, social media, and smartphone apps. Citizen science is a special type of crowdsourcing, where the participants contribute to or collect data for scientific research projects. Citizen science benefits both the researchers and the people engaged in it. Researchers can collect data that they otherwise would not be able to while the participants learn about the subject they are engaged with. For example, when using iNaturalist, an app that anyone can download on their phone, people collect data and learn about plant or animal species \cite{van2018inaturalist}. Increasingly, citizen science platforms are going mobile with emerging technologies and shifting paradigms \cite{newman2012future}. To effectively engage in citizen science projects, the participants often need to develop skills for detecting, identifying, and annotating phenomena or entities from visual inputs. At present, it is expected that the participants already have these skill sets, or they can quickly develop these by following a set of instructions or tutorials \cite{rosser2018tutorial}. However, for novice participants, it is often not as easy to understand some new phenomena or entities in real life just by following a set of instructions or tutorials. In this article, we use the conventional term "researchers" to describe the group that runs the research projects and need to collect data, and "participants" to describe the group that collects the data and contribute to the project using a citizen science platform or app \cite{eitzel2017citizen}.

\begin{figure*}[ht]
  \centering
  \includegraphics[width=0.95\linewidth]{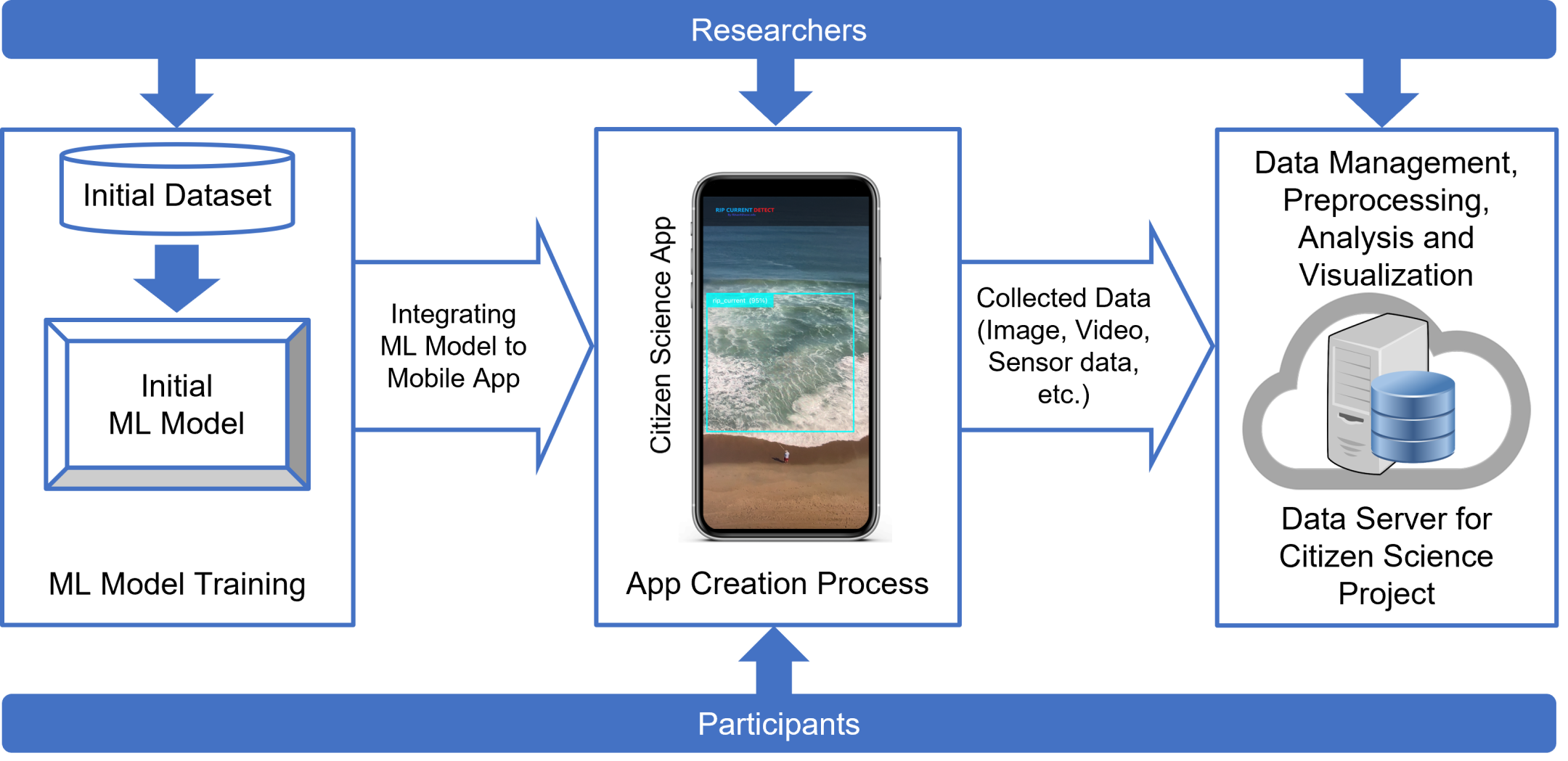}
  \caption{Overview of the components of our open-source citizen science platform architecture.}
  \Description{This figure shows an overview of the components of our open-source citizen science platform architecture.}
  \label{fig:overview}
\end{figure*}

To illustrate the challenges faced by potential participants in collecting research quality data, we describe the rip current detection problem \cite{philip2016detecting}. Rip currents are safety hazards that can claim human lives. To answer questions like: "Are there rip currents at this beach?" the researcher needs to gather data that an army of participants can conveniently collect. However, spotting rip currents can be challenging for novice participants unless they are familiar with this subject matter \cite{brannstrom2015you}. Recent works demonstrated that rip currents can be detected using ML approaches \cite{de2021automated,maryan2019machine}. Providing real-time ML-based guidance using bounding boxes around rip currents in the live camera feed of the mobile app enables the participants to learn to spot rip currents and collect data more effectively. This facilitates the effective engagement of people who may have less familiarity with rip currents. The ML-based guidance can work as an educational tool for the participants as well, alerting them of potential danger. While this example focuses on rip currents, it can be replaced with a wide variety of fields and natural phenomena on which researchers are interested in collecting data, such as biological sciences, aquaculture, geomorphology, drought, and flooding indicators, to name a few. In all these cases, mobile apps with real-time ML-based guidance systems enable the participants to recognize and collect the correct data.

Indeed, there are similar infrastructures that allow one to create people-powered apps like those provided by Zooniverse, SPOTTERON, Anecdata, etc. \cite{liu2021citizen}. However, these general-purpose citizen science app builders are limited to providing standardized purpose-specific tools and features, such as task assignment and data uploading for crowdsourced research projects. Also, the data collection processes entirely rely on human skills as there is no integrated ML support in the client apps. Some apps like iNaturalist have server-side ML capabilities \cite{van2018inaturalist}. However, to use server-side ML in real-time, continuous high-speed connectivity is required, which can be expensive, and internet connectivity is not available in many remote places where some projects might need to collect data. While considering building ML capabilities on top of the architectures of existing open-source systems (iNaturalist, Zooniverse, etc.), our analysis showed that they are not designed to integrate with client-side ML. It is possible to develop a new citizen science app from scratch with ML support for each project. However, developing and deploying each of these individual apps would take months, if not years. For example, an app with client-side ML capabilities to classify plant and animal species is Seek by iNaturalist \cite{van2018inaturalist}. While iNaturalist already has a well-developed app with server-side ML, they had to create Seek from scratch to add client app-side ML support. While the topics of different citizen science projects may seem far afield from each other, common needs for collecting visual data tie these domain problems together. So, ML-powered apps created using a general-purpose citizen science platform can help participants recognize the correct visual data and increase the efficiency of the data collection process in a wide variety of research domains.

This article introduces an open-source software platform that allows a domain researcher to quickly create citizen science apps with integrated ML models to collect visual data, even if they don't have a computer science background. Existing ML-powered citizen science apps often involve development stages that take months or years to deploy. Our proposed platform reduces many of those stages by providing a common feature set under a single framework shared by all apps. This will enable rapid prototyping and faster deployments allowing researchers without a large budget or projects that are more investigative than long-term in nature to engage in productive work.

\section{Related Works}

\begin{figure*}[ht]
  \centering
  \includegraphics[width=0.9\linewidth]{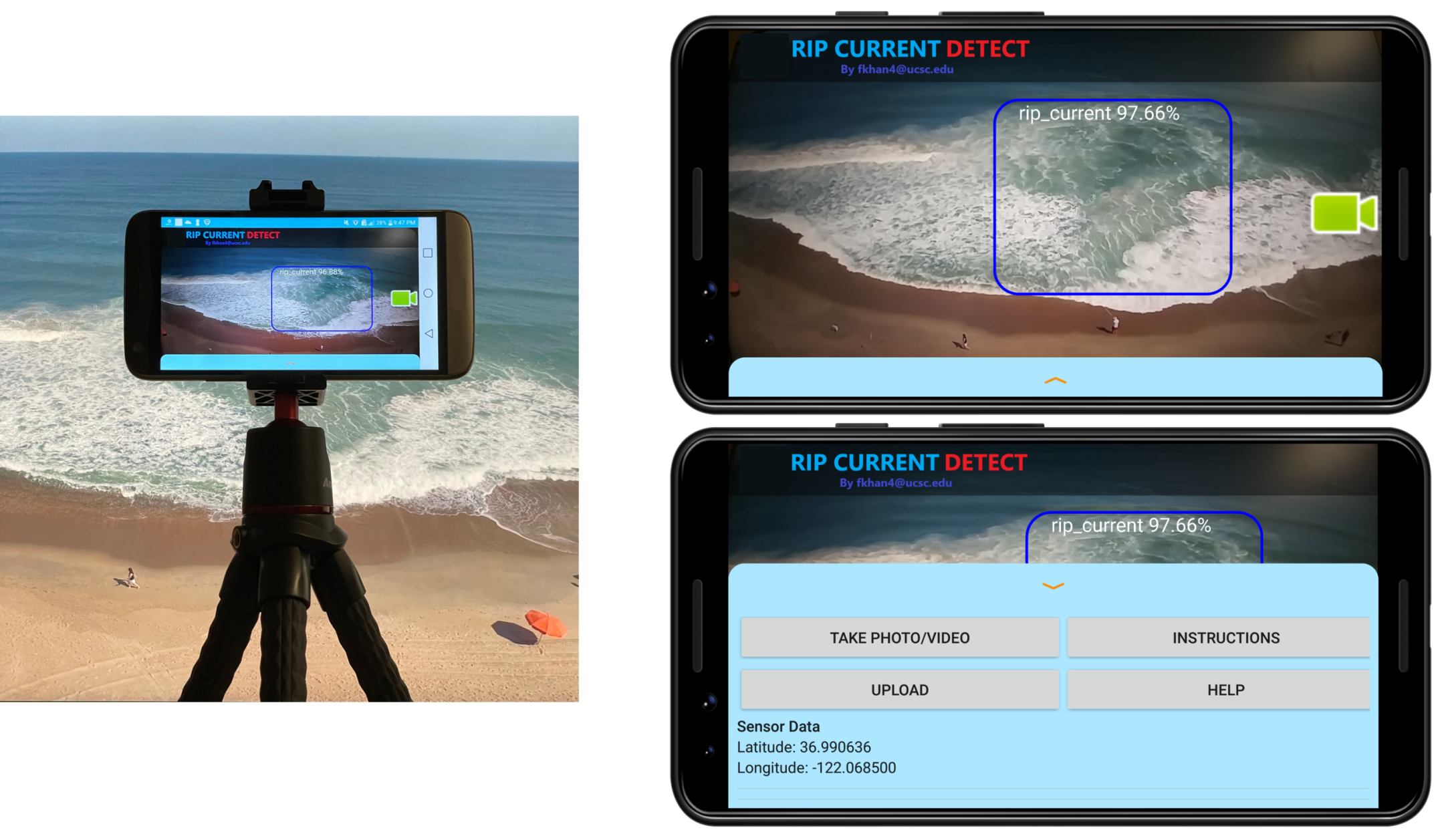}
  \caption{ Left: app created by our citizen science platform in field work. Right: (top) detection bounding box with confidence score and (bottom) graphical user interface showing sensor information, data uploading and other options.}
  \Description{This figure shows an app created by our citizen science platform in field work.}
  \label{fig:implementation}
\end{figure*}

We studied the most popular citizen science app creation platforms that allow one to create people-powered mobile apps. Zooniverse is a free citizen science web portal that allows creating projects for different domains \cite{barber2018zooniverse}. A sample project from Zooniverse is \href{https://www.zooniverse.org/projects/benjamin-dot-richards/oceaneyes}{OceanEyes}, where volunteers are sought to help count and label the fishes in the images that the researchers' cameras have collected.  It illustrates how having an ML model to identify the fish species can be highly beneficial. Anecdata is another free online citizen science platform that has similar features as Zooniverse \cite{disney2018anecdata}. Another fully mobile app-based citizen science platform is SPOTTERON \cite{liu2021citizen}. All the citizen science apps using this framework have the same look and have an easy-to-customize GUI for various projects. Powered by SPOTTERON, another citizen science app is \href{https://www.coastsnap.com/}{CoastSnap}, which uses uploaded beach photos to understand how coastlines might change in the coming decades \cite{harley2018coastsnap}. App Movement is another authoring platform for community-created mobile apps, which provides automatic development and deployment of the app by customizing a common template \cite{garbett2016app}. However, the client-side ML component is not available on any of these platforms.

\section{System Components of Authoring Platform}

The main components of our proposed citizen science platform are the mobile app (middle), the ML models in the app (left), and the server-side components (right) shown in Fig. \ref{fig:overview}. The app contains the ML model and provides the primary interface for the participants. It provides standardized purpose-specific tools and features for crowdsourced research projects. Built-in tools include instructions, tutorials, data saving, uploading, etc. Since we aim to facilitate visual data collection, the app includes a camera tool with a live view that doubles as the visualizer for the ML model (e.g., bounding boxes around the detected objects). New projects initially start with a blank template with these built-in features, functionalities, and default look-and-feel that can be customized later. 

For each project, the ML model needs to be trained with an initial dataset. We assume that the researcher has this. If a project has no data at all, the researcher's team will need to collect some limited initial training data to create a rudimentary model that can be improved via continual learning as more data is collected \cite{schwarz2018progress}. We also assume the researchers themselves and other domain experts will use the app as "expert participants" who can collect higher quality data and provide labels that correct the misclassifications or false positive detections from the rudimentary model. Thus, there is an opportunity to engage the "expert participants" more intimately by being part of the process to improve the ML model through confirmation or refutation. If a "perfect" training dataset exists for a project, the researcher can directly use that for training the model and quickly start large-scale deployment for data collection.

The trained model is integrated with the app before building and deploying the app. When the participants use the app for data collection, the ML model runs with the app and guides with classifications or annotations (e.g., bounding boxes) to help them recognize the object for data collection. The models are fully compatible with mobile device architectures and run locally without internet connectivity and back-end server support. The back-end primarily works as the repositories for collecting the data that the participants upload. Other optional features included on the server-side are a companion website, user account, data management, data explorer, analysis and visualization, server-side ML apps with ML models requiring more computational power than mobile devices, etc. There is a primary server for managing all the apps for each citizen science project in our architecture. However, each project has its own data storage server (physical or cloud) for storing the collected datasets.

\begin{figure*}[ht]
  \centering
  \includegraphics[width=1.0\linewidth]{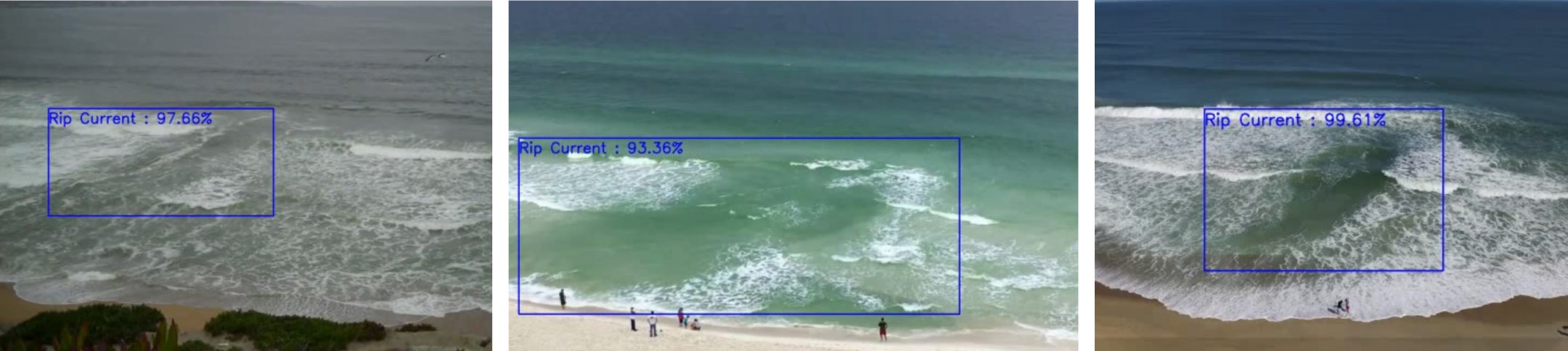}
  \caption{Some examples from our test cases of real-time rip current detection and data collection using our mobile app are illustrated here. The location of the rip currents is visualized using the blue bounding box with the label and the confidence score of detection.}
  \Description{This figure shows some examples from our test cases of real-time rip current detection and data collection using our mobile app.}
  \label{fig:rip_detect}
\end{figure*}

\section{Implementation}

The architecture of our citizen science platform is a standard client-server system (Fig. \ref{fig:overview}). The app and the integrated ML model run on the client devices, e.g., smartphones. The phone camera provides real-time visual inputs for the ML model to process (Fig. \ref{fig:implementation}). We use ML models based on TensorFlow Lite \cite{tensorflow2015-whitepaper}. These models are small and optimized to run on limited computational resources on mobile devices. The same trained model runs on both Android and iOS versions of the app. At present, single-shot detector (SSD) models are supported by TensorFlow Lite. For our test cases, we used SSD MobileNetV2 \cite{sandler2018mobilenetv2} and EfficientDet \cite{tan2020efficientdet}, and trained the models using transfer learning \cite{alsing2018mobile}. 

\begin{figure*}[ht]
  \centering
  \includegraphics[width=1.0\linewidth]{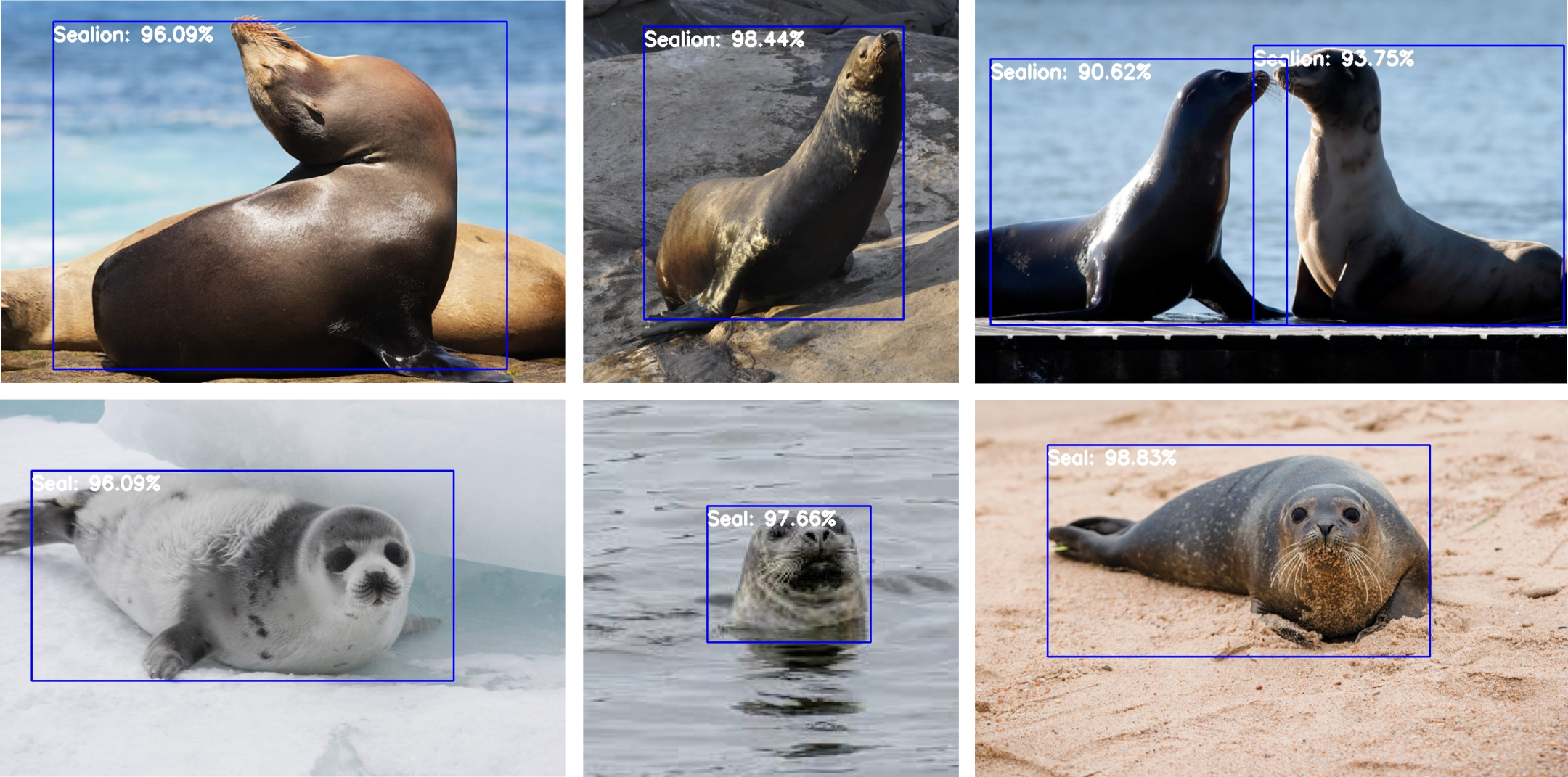}
  \caption{Some examples from our test cases of real-time detection and differentiation between sea lions and seals using our mobile app are illustrated here. The three images on the top row show one or more sea lions detected and visualized using separate bounding boxes with labels and confidence scores. Similarly, the three images on the bottom row show detected seals in different environments.}
    \Description{This figure shows some examples from our test cases of real-time detection and differentiation between sea lions and seals using our mobile app.}
  \label{fig:seal_detect}
\end{figure*}

Our platform simplifies the authoring process using a set of internal and external web-based tools. The fully guided app creation project starts on the website of our authoring platform. The website has the instructions and tools for the project creators to train a model using their custom dataset to bootstrap their project. Then, we provide another web-based tool to integrate the trained model and compile the app. Once the app is ready, it can be uploaded to the app distribution services for mobile devices to make it available for the participants to download and use.
 
Based on the guidance and feedback from the ML model running in the app, the participants can decide which data they want to capture. As the ML model integrated into the app runs locally on mobile devices, no server support or internet connectivity is required. The data is captured and initially stored in the local storage of the smartphone. The participants can later select the data they want to upload to the server. A larger ML model on the server-side can be used to analyze further and verify the collected data utilizing more powerful computational resources.

\section{Results and Discussion}

We did some initial testing of this architecture on two separate citizen science projects. The results from the two use cases are presented below.

\subsection{Use Case 1: Rip Current Detection for Beach Safety}

Our first use case is a citizen science app to collect data on rip current events for use in rip forecast model verification and creating a database for rip current research \cite{dusek2013probabilistic}. The ML models for rip current detection such as the one reported in \cite{de2021automated} are too large and computational resource-intensive for mobile deployment. Using the authoring platform described above, we created a mobile app that would contribute to beach safety by alerting people to the presence and location of rip currents, if any (Fig. \ref{fig:rip_detect}). The mobile-optimized ML model in the app helps the participants with no previous experience to spot rip currents and collect data for the citizen science project. We bootstrapped the training process by using the data from \cite{de2021automated}.  Also, the labels of data provided by "expert participants" who are more familiar with rip currents (e.g., lifeguards, local surfers, etc.) can improve data quality.  This app is currently being prepared for an upcoming pilot study this summer to be conducted at various locations in the US. This pilot study will allow us to improve the app further before it goes "live".

\subsection{Use Case 2: Biodiversity Analysis}

Biodiversity analysis is important for many research groups, such as those with a focus on biological science, aquaculture, marine biology, etc. Researchers may need to collect data about some endangered species; other times, they need data to analyze the biodiversity in some specific area \cite{willi2019identifying, wood2021accuracy}. In this use case, we trained a model with images of sea lions and seals to demonstrate our app's usability for these types of research projects. Many sea lion species are considered as endangered \cite{chilvers2017conservation}, and collecting data about them are needed for marine biology research and conservation groups \cite{brown2020california}. However, it can be difficult for novice participants to differentiate between seals and sea lions \cite{wood2021accuracy}. Using our ML-powered app, the participants can detect and differentiate these two species (Fig. \ref{fig:seal_detect}). With further training data and continual learning, this app can be modified to detect and differentiate among various sub-species \cite{hann2018obstacles}.

\section{Conclusion and Future Works}

This article presents an overview of an open-source platform for creating client-side ML-powered citizen science apps to improve data collection quality and efficiency. We demonstrate the use of the authoring platform with two real-world examples. As we continue working on the citizen science platform, we plan to optimize the overall process, including an enhanced user interface for mobile apps, support for a wider variety of ML models, and more server-side services.

\begin{acks}
This report was prepared in part as a result of work sponsored by the Southeast Coastal Ocean Observing Regional Association (SECOORA) with National Oceanic and Atmospheric Administration (NOAA) financial assistance award number NA20NOS0120220. 

The scientific results and conclusions, as well as any views or opinions expressed herein, are those of the author(s) and do not necessarily reflect the views of NOAA or the Department of Commerce.
\end{acks}

\bibliographystyle{ACM-Reference-Format}
\balance
\bibliography{main}


\begin{thebibliography}{24}

%%% the Web address.

\ifx \showCODEN    \undefined \def \showCODEN     #1{\unskip}     \fi
\ifx \showDOI      \undefined \def \showDOI       #1{#1}\fi
\ifx \showISBNx    \undefined \def \showISBNx     #1{\unskip}     \fi
\ifx \showISBNxiii \undefined \def \showISBNxiii  #1{\unskip}     \fi
\ifx \showISSN     \undefined \def \showISSN      #1{\unskip}     \fi
\ifx \showLCCN     \undefined \def \showLCCN      #1{\unskip}     \fi
\ifx \shownote     \undefined \def \shownote      #1{#1}          \fi
\ifx \showarticletitle \undefined \def \showarticletitle #1{#1}   \fi
\ifx \showURL      \undefined \def \showURL       {\relax}        \fi
% invisible to TeX
\providecommand\bibfield[2]{#2}
\providecommand\bibinfo[2]{#2}
\providecommand\natexlab[1]{#1}
\providecommand\showeprint[2][]{arXiv:#2}

\bibitem[\protect\citeauthoryear{Abadi, Agarwal, Barham, Brevdo, Chen, Citro,
  Corrado, Davis, Dean, Devin, Ghemawat, Goodfellow, Harp, Irving, Isard, Jia,
  Jozefowicz, Kaiser, Kudlur, Levenberg, Man\'{e}, Monga, Moore, Murray, Olah,
  Schuster, Shlens, Steiner, Sutskever, Talwar, Tucker, Vanhoucke, Vasudevan,
  Vi\'{e}gas, Vinyals, Warden, Wattenberg, Wicke, Yu, and Zheng}{Abadi
  et~al\mbox{.}}{2015}]%
        {tensorflow2015-whitepaper}
\bibfield{author}{\bibinfo{person}{Mart\'{\i}n Abadi}, \bibinfo{person}{Ashish
  Agarwal}, \bibinfo{person}{Paul Barham}, \bibinfo{person}{Eugene Brevdo},
  \bibinfo{person}{Zhifeng Chen}, \bibinfo{person}{Craig Citro},
  \bibinfo{person}{Greg~S. Corrado}, \bibinfo{person}{Andy Davis},
  \bibinfo{person}{Jeffrey Dean}, \bibinfo{person}{Matthieu Devin},
  \bibinfo{person}{Sanjay Ghemawat}, \bibinfo{person}{Ian Goodfellow},
  \bibinfo{person}{Andrew Harp}, \bibinfo{person}{Geoffrey Irving},
  \bibinfo{person}{Michael Isard}, \bibinfo{person}{Yangqing Jia},
  \bibinfo{person}{Rafal Jozefowicz}, \bibinfo{person}{Lukasz Kaiser},
  \bibinfo{person}{Manjunath Kudlur}, \bibinfo{person}{Josh Levenberg},
  \bibinfo{person}{Dandelion Man\'{e}}, \bibinfo{person}{Rajat Monga},
  \bibinfo{person}{Sherry Moore}, \bibinfo{person}{Derek Murray},
  \bibinfo{person}{Chris Olah}, \bibinfo{person}{Mike Schuster},
  \bibinfo{person}{Jonathon Shlens}, \bibinfo{person}{Benoit Steiner},
  \bibinfo{person}{Ilya Sutskever}, \bibinfo{person}{Kunal Talwar},
  \bibinfo{person}{Paul Tucker}, \bibinfo{person}{Vincent Vanhoucke},
  \bibinfo{person}{Vijay Vasudevan}, \bibinfo{person}{Fernanda Vi\'{e}gas},
  \bibinfo{person}{Oriol Vinyals}, \bibinfo{person}{Pete Warden},
  \bibinfo{person}{Martin Wattenberg}, \bibinfo{person}{Martin Wicke},
  \bibinfo{person}{Yuan Yu}, {and} \bibinfo{person}{Xiaoqiang Zheng}.}
  \bibinfo{year}{2015}\natexlab{}.
\newblock \bibinfo{title}{{TensorFlow}: Large-Scale Machine Learning on
  Heterogeneous Systems}.
\newblock
\newblock
\urldef\tempurl%
\url{https://www.tensorflow.org/}
\showURL{%
\tempurl}
\newblock
\shownote{Software available from tensorflow.org.}


\bibitem[\protect\citeauthoryear{Alsing}{Alsing}{2018}]%
        {alsing2018mobile}
\bibfield{author}{\bibinfo{person}{Oscar Alsing}.}
  \bibinfo{year}{2018}\natexlab{}.
\newblock \emph{\bibinfo{title}{Mobile Object Detection using TensorFlow Lite
  and Transfer Learning}}.
\newblock \bibinfo{thesistype}{Master's\ thesis}. \bibinfo{school}{KTH, School
  of Electrical Engineering and Computer Science (EECS)}.
\newblock
\urldef\tempurl%
\url{https://www.diva-portal.org/smash/record.jsf?pid=diva2:1242627}
\showURL{%
\tempurl}


\bibitem[\protect\citeauthoryear{Barber}{Barber}{2018}]%
        {barber2018zooniverse}
\bibfield{author}{\bibinfo{person}{Samuel~T Barber}.}
  \bibinfo{year}{2018}\natexlab{}.
\newblock \showarticletitle{The zooniverse is expanding: crowdsourced solutions
  to the hidden collections problem and the rise of the revolutionary
  cataloging interface}.
\newblock \bibinfo{journal}{\emph{Journal of Library Metadata}}
  \bibinfo{volume}{18}, \bibinfo{number}{2} (\bibinfo{year}{2018}),
  \bibinfo{pages}{85--111}.
\newblock


\bibitem[\protect\citeauthoryear{Brannstrom, Brown, Houser, Trimble, and
  Santos}{Brannstrom et~al\mbox{.}}{2015}]%
        {brannstrom2015you}
\bibfield{author}{\bibinfo{person}{Christian Brannstrom},
  \bibinfo{person}{Heather~Lee Brown}, \bibinfo{person}{Chris Houser},
  \bibinfo{person}{Sarah Trimble}, {and} \bibinfo{person}{Anna Santos}.}
  \bibinfo{year}{2015}\natexlab{}.
\newblock \showarticletitle{“You can't see them from sitting here”:
  Evaluating beach user understanding of a rip current warning sign}.
\newblock \bibinfo{journal}{\emph{Applied Geography}}  \bibinfo{volume}{56}
  (\bibinfo{year}{2015}), \bibinfo{pages}{61--70}.
\newblock


\bibitem[\protect\citeauthoryear{Brown, Wright, Tennis, and Jeffries}{Brown
  et~al\mbox{.}}{2020}]%
        {brown2020california}
\bibfield{author}{\bibinfo{person}{Robin~F Brown}, \bibinfo{person}{Bryan~E
  Wright}, \bibinfo{person}{Matthew~J Tennis}, {and} \bibinfo{person}{Steven
  Jeffries}.} \bibinfo{year}{2020}\natexlab{}.
\newblock \showarticletitle{California sea lion (Zalophus californianus)
  monitoring in the Lower Columbia River, 1997--2018}.
\newblock \bibinfo{journal}{\emph{Northwestern Naturalist}}
  \bibinfo{volume}{101}, \bibinfo{number}{2} (\bibinfo{year}{2020}),
  \bibinfo{pages}{92--103}.
\newblock


\bibitem[\protect\citeauthoryear{Chilvers and Meyer}{Chilvers and
  Meyer}{2017}]%
        {chilvers2017conservation}
\bibfield{author}{\bibinfo{person}{B~Louise Chilvers} {and}
  \bibinfo{person}{Stefan Meyer}.} \bibinfo{year}{2017}\natexlab{}.
\newblock \showarticletitle{Conservation needs for the endangered New Zealand
  sea lion, Phocarctos hookeri}.
\newblock \bibinfo{journal}{\emph{Aquatic Conservation: Marine and Freshwater
  Ecosystems}} \bibinfo{volume}{27}, \bibinfo{number}{4}
  (\bibinfo{year}{2017}), \bibinfo{pages}{846--855}.
\newblock


\bibitem[\protect\citeauthoryear{de~Silva, Mori, Dusek, Davis, and
  Pang}{de~Silva et~al\mbox{.}}{2021}]%
        {de2021automated}
\bibfield{author}{\bibinfo{person}{Akila de Silva}, \bibinfo{person}{Issei
  Mori}, \bibinfo{person}{Gregory Dusek}, \bibinfo{person}{James Davis}, {and}
  \bibinfo{person}{Alex Pang}.} \bibinfo{year}{2021}\natexlab{}.
\newblock \showarticletitle{Automated rip current detection with region based
  convolutional neural networks}.
\newblock \bibinfo{journal}{\emph{Coastal Engineering}}  \bibinfo{volume}{166}
  (\bibinfo{year}{2021}), \bibinfo{pages}{103859}.
\newblock


\bibitem[\protect\citeauthoryear{Disney, Bailey, Farrell, Taylor, and
  McGreavy}{Disney et~al\mbox{.}}{2018}]%
        {disney2018anecdata}
\bibfield{author}{\bibinfo{person}{Jane Disney}, \bibinfo{person}{Duncan
  Bailey}, \bibinfo{person}{Anna Farrell}, \bibinfo{person}{Ashley Taylor},
  {and} \bibinfo{person}{Bridie McGreavy}.} \bibinfo{year}{2018}\natexlab{}.
\newblock \showarticletitle{Anecdata. org: An online citizen science platform
  for Building Climate Resilient Communities}. In
  \bibinfo{booktitle}{\emph{OCEANS 2018 MTS/IEEE Charleston}}. IEEE,
  \bibinfo{publisher}{IEEE}, \bibinfo{address}{Charleston, SC, USA},
  \bibinfo{pages}{1--4}.
\newblock


\bibitem[\protect\citeauthoryear{Dusek and Seim}{Dusek and Seim}{2013}]%
        {dusek2013probabilistic}
\bibfield{author}{\bibinfo{person}{G Dusek} {and} \bibinfo{person}{H Seim}.}
  \bibinfo{year}{2013}\natexlab{}.
\newblock \showarticletitle{A probabilistic rip current forecast model}.
\newblock \bibinfo{journal}{\emph{Journal of Coastal Research}}
  \bibinfo{volume}{29}, \bibinfo{number}{4} (\bibinfo{year}{2013}),
  \bibinfo{pages}{909--925}.
\newblock


\bibitem[\protect\citeauthoryear{Eitzel, Cappadonna, Santos-Lang, Duerr,
  Virapongse, West, Kyba, Bowser, Cooper, Sforzi, et~al\mbox{.}}{Eitzel
  et~al\mbox{.}}{2017}]%
        {eitzel2017citizen}
\bibfield{author}{\bibinfo{person}{Melissa~V Eitzel},
  \bibinfo{person}{Jessica~L Cappadonna}, \bibinfo{person}{Chris Santos-Lang},
  \bibinfo{person}{Ruth~Ellen Duerr}, \bibinfo{person}{Arika Virapongse},
  \bibinfo{person}{Sarah~Elizabeth West}, \bibinfo{person}{Christopher Kyba},
  \bibinfo{person}{Anne Bowser}, \bibinfo{person}{Caren~Beth Cooper},
  \bibinfo{person}{Andrea Sforzi}, {et~al\mbox{.}}}
  \bibinfo{year}{2017}\natexlab{}.
\newblock \showarticletitle{Citizen science terminology matters: Exploring key
  terms}.
\newblock \bibinfo{journal}{\emph{Citizen Science: Theory and Practice}}
  \bibinfo{volume}{2}, \bibinfo{number}{1} (\bibinfo{year}{2017}),
  \bibinfo{pages}{1–20}.
\newblock


\bibitem[\protect\citeauthoryear{Garbett, Comber, Jenkins, and Olivier}{Garbett
  et~al\mbox{.}}{2016}]%
        {garbett2016app}
\bibfield{author}{\bibinfo{person}{Andrew Garbett}, \bibinfo{person}{Rob
  Comber}, \bibinfo{person}{Edward Jenkins}, {and} \bibinfo{person}{Patrick
  Olivier}.} \bibinfo{year}{2016}\natexlab{}.
\newblock \bibinfo{booktitle}{\emph{App Movement: A Platform for Community
  Commissioning of Mobile Applications}}.
\newblock \bibinfo{publisher}{Association for Computing Machinery},
  \bibinfo{address}{New York, NY, USA}, \bibinfo{pages}{26–37}.
\newblock
\showISBNx{9781450333627}
\urldef\tempurl%
\url{https://doi.org/10.1145/2858036.2858094}
\showURL{%
\tempurl}


\bibitem[\protect\citeauthoryear{Hann, Stelle, Szabo, and Torres}{Hann
  et~al\mbox{.}}{2018}]%
        {hann2018obstacles}
\bibfield{author}{\bibinfo{person}{Courtney~H Hann}, \bibinfo{person}{Lei~Lani
  Stelle}, \bibinfo{person}{Andrew Szabo}, {and} \bibinfo{person}{Leigh~G
  Torres}.} \bibinfo{year}{2018}\natexlab{}.
\newblock \showarticletitle{Obstacles and opportunities of using a mobile app
  for marine mammal research}.
\newblock \bibinfo{journal}{\emph{ISPRS International Journal of
  Geo-Information}} \bibinfo{volume}{7}, \bibinfo{number}{5}
  (\bibinfo{year}{2018}), \bibinfo{pages}{169}.
\newblock


\bibitem[\protect\citeauthoryear{Harley, Kinsela, S{\'a}nchez-Garc{\'\i}a, and
  Vos}{Harley et~al\mbox{.}}{2018}]%
        {harley2018coastsnap}
\bibfield{author}{\bibinfo{person}{Mitchell Harley}, \bibinfo{person}{Michael
  Kinsela}, \bibinfo{person}{Elena~S{\'a}nchez S{\'a}nchez-Garc{\'\i}a}, {and}
  \bibinfo{person}{Kilian Vos}.} \bibinfo{year}{2018}\natexlab{}.
\newblock \showarticletitle{CoastSnap: Crowd-Sourced Shoreline Change Mapping
  using Smartphones}. In \bibinfo{booktitle}{\emph{AGU Fall Meeting
  Abstracts}}, Vol.~\bibinfo{volume}{2018}. \bibinfo{publisher}{SAO/NASA
  Astrophysics Data System}, \bibinfo{address}{USA},
  \bibinfo{pages}{EP52D--26}.
\newblock


\bibitem[\protect\citeauthoryear{Horn, Aodha, Song, Cui, Sun, Shepard, Adam,
  Perona, and Belongie}{Horn et~al\mbox{.}}{2018}]%
        {van2018inaturalist}
\bibfield{author}{\bibinfo{person}{G.~Van Horn}, \bibinfo{person}{O.~Mac
  Aodha}, \bibinfo{person}{Y. Song}, \bibinfo{person}{Y. Cui},
  \bibinfo{person}{C. Sun}, \bibinfo{person}{A. Shepard}, \bibinfo{person}{H.
  Adam}, \bibinfo{person}{P. Perona}, {and} \bibinfo{person}{S. Belongie}.}
  \bibinfo{year}{2018}\natexlab{}.
\newblock \showarticletitle{The iNaturalist Species Classification and
  Detection Dataset}. In \bibinfo{booktitle}{\emph{2018 IEEE/CVF Conference on
  Computer Vision and Pattern Recognition (CVPR)}}. \bibinfo{publisher}{IEEE
  Computer Society}, \bibinfo{address}{Los Alamitos, CA, USA},
  \bibinfo{pages}{8769--8778}.
\newblock
\urldef\tempurl%
\url{https://doi.org/10.1109/CVPR.2018.00914}
\showDOI{\tempurl}


\bibitem[\protect\citeauthoryear{Liu, D{\"o}rler, Heigl, and Grossberndt}{Liu
  et~al\mbox{.}}{2021}]%
        {liu2021citizen}
\bibfield{author}{\bibinfo{person}{Hai-Ying Liu}, \bibinfo{person}{Daniel
  D{\"o}rler}, \bibinfo{person}{Florian Heigl}, {and} \bibinfo{person}{Sonja
  Grossberndt}.} \bibinfo{year}{2021}\natexlab{}.
\newblock \showarticletitle{Citizen Science Platforms}.
\newblock In \bibinfo{booktitle}{\emph{The Science of Citizen Science}}.
  \bibinfo{publisher}{Springer}, \bibinfo{address}{Cham, Switzerland},
  \bibinfo{pages}{439--459}.
\newblock


\bibitem[\protect\citeauthoryear{Maryan, Hoque, Michael, Ioup, and
  Abdelguerfi}{Maryan et~al\mbox{.}}{2019}]%
        {maryan2019machine}
\bibfield{author}{\bibinfo{person}{Corey Maryan}, \bibinfo{person}{Md~Tamjidul
  Hoque}, \bibinfo{person}{Christopher Michael}, \bibinfo{person}{Elias Ioup},
  {and} \bibinfo{person}{Mahdi Abdelguerfi}.} \bibinfo{year}{2019}\natexlab{}.
\newblock \showarticletitle{Machine learning applications in detecting rip
  channels from images}.
\newblock \bibinfo{journal}{\emph{Applied Soft Computing}}
  \bibinfo{volume}{78} (\bibinfo{year}{2019}), \bibinfo{pages}{84--93}.
\newblock


\bibitem[\protect\citeauthoryear{Newman, Wiggins, Crall, Graham, Newman, and
  Crowston}{Newman et~al\mbox{.}}{2012}]%
        {newman2012future}
\bibfield{author}{\bibinfo{person}{Greg Newman}, \bibinfo{person}{Andrea
  Wiggins}, \bibinfo{person}{Alycia Crall}, \bibinfo{person}{Eric Graham},
  \bibinfo{person}{Sarah Newman}, {and} \bibinfo{person}{Kevin Crowston}.}
  \bibinfo{year}{2012}\natexlab{}.
\newblock \showarticletitle{The future of citizen science: emerging
  technologies and shifting paradigms}.
\newblock \bibinfo{journal}{\emph{Frontiers in Ecology and the Environment}}
  \bibinfo{volume}{10}, \bibinfo{number}{6} (\bibinfo{year}{2012}),
  \bibinfo{pages}{298--304}.
\newblock


\bibitem[\protect\citeauthoryear{Philip and Pang}{Philip and Pang}{2016}]%
        {philip2016detecting}
\bibfield{author}{\bibinfo{person}{S. Philip} {and} \bibinfo{person}{A. Pang}.}
  \bibinfo{year}{2016}\natexlab{}.
\newblock \showarticletitle{Detecting and Visualizing Rip Current Using Optical
  Flow}. In \bibinfo{booktitle}{\emph{Proceedings of the Eurographics / IEEE
  VGTC Conference on Visualization: Short Papers}} (Groningen, The Netherlands)
  \emph{(\bibinfo{series}{EuroVis '16})}. \bibinfo{publisher}{Eurographics
  Association}, \bibinfo{address}{Goslar, DEU}, \bibinfo{pages}{19–23}.
\newblock


\bibitem[\protect\citeauthoryear{Rosser and Wiggins}{Rosser and
  Wiggins}{2018}]%
        {rosser2018tutorial}
\bibfield{author}{\bibinfo{person}{Holly Rosser} {and} \bibinfo{person}{Andrea
  Wiggins}.} \bibinfo{year}{2018}\natexlab{}.
\newblock \showarticletitle{Tutorial Designs and Task Types in Zooniverse}. In
  \bibinfo{booktitle}{\emph{Companion of the 2018 ACM Conference on Computer
  Supported Cooperative Work and Social Computing}} (Jersey City, NJ, USA)
  \emph{(\bibinfo{series}{CSCW '18})}. \bibinfo{publisher}{Association for
  Computing Machinery}, \bibinfo{address}{New York, NY, USA},
  \bibinfo{pages}{177–180}.
\newblock
\showISBNx{9781450360180}
\urldef\tempurl%
\url{https://doi.org/10.1145/3272973.3274049}
\showDOI{\tempurl}


\bibitem[\protect\citeauthoryear{Sandler, Howard, Zhu, Zhmoginov, and
  Chen}{Sandler et~al\mbox{.}}{2018}]%
        {sandler2018mobilenetv2}
\bibfield{author}{\bibinfo{person}{Mark Sandler}, \bibinfo{person}{Andrew
  Howard}, \bibinfo{person}{Menglong Zhu}, \bibinfo{person}{Andrey Zhmoginov},
  {and} \bibinfo{person}{Liang-Chieh Chen}.} \bibinfo{year}{2018}\natexlab{}.
\newblock \showarticletitle{MobileNetV2: Inverted Residuals and Linear
  Bottlenecks}. In \bibinfo{booktitle}{\emph{2018 IEEE/CVF Conference on
  Computer Vision and Pattern Recognition}}. \bibinfo{publisher}{IEEE Computer
  Society}, \bibinfo{address}{Los Alamitos, CA, USA},
  \bibinfo{pages}{4510--4520}.
\newblock
\urldef\tempurl%
\url{https://doi.org/10.1109/CVPR.2018.00474}
\showDOI{\tempurl}


\bibitem[\protect\citeauthoryear{Schwarz, Czarnecki, Luketina,
  Grabska-Barwinska, Teh, Pascanu, and Hadsell}{Schwarz et~al\mbox{.}}{2018}]%
        {schwarz2018progress}
\bibfield{author}{\bibinfo{person}{Jonathan Schwarz}, \bibinfo{person}{Wojciech
  Czarnecki}, \bibinfo{person}{Jelena Luketina}, \bibinfo{person}{Agnieszka
  Grabska-Barwinska}, \bibinfo{person}{Yee~Whye Teh}, \bibinfo{person}{Razvan
  Pascanu}, {and} \bibinfo{person}{Raia Hadsell}.}
  \bibinfo{year}{2018}\natexlab{}.
\newblock \showarticletitle{Progress \& compress: A scalable framework for
  continual learning}. In \bibinfo{booktitle}{\emph{International Conference on
  Machine Learning}}. PMLR, \bibinfo{publisher}{PMLR}, \bibinfo{address}{USA},
  \bibinfo{pages}{4528--4537}.
\newblock


\bibitem[\protect\citeauthoryear{Tan, Pang, and Le}{Tan et~al\mbox{.}}{2020}]%
        {tan2020efficientdet}
\bibfield{author}{\bibinfo{person}{M. Tan}, \bibinfo{person}{R. Pang}, {and}
  \bibinfo{person}{Q.~V. Le}.} \bibinfo{year}{2020}\natexlab{}.
\newblock \showarticletitle{EfficientDet: Scalable and Efficient Object
  Detection}. In \bibinfo{booktitle}{\emph{2020 IEEE/CVF Conference on Computer
  Vision and Pattern Recognition (CVPR)}}. \bibinfo{publisher}{IEEE Computer
  Society}, \bibinfo{address}{Los Alamitos, CA, USA},
  \bibinfo{pages}{10778--10787}.
\newblock
\urldef\tempurl%
\url{https://doi.org/10.1109/CVPR42600.2020.01079}
\showDOI{\tempurl}


\bibitem[\protect\citeauthoryear{Willi, Pitman, Cardoso, Locke, Swanson, Boyer,
  Veldthuis, and Fortson}{Willi et~al\mbox{.}}{2019}]%
        {willi2019identifying}
\bibfield{author}{\bibinfo{person}{Marco Willi}, \bibinfo{person}{Ross~T
  Pitman}, \bibinfo{person}{Anabelle~W Cardoso}, \bibinfo{person}{Christina
  Locke}, \bibinfo{person}{Alexandra Swanson}, \bibinfo{person}{Amy Boyer},
  \bibinfo{person}{Marten Veldthuis}, {and} \bibinfo{person}{Lucy Fortson}.}
  \bibinfo{year}{2019}\natexlab{}.
\newblock \showarticletitle{Identifying animal species in camera trap images
  using deep learning and citizen science}.
\newblock \bibinfo{journal}{\emph{Methods in Ecology and Evolution}}
  \bibinfo{volume}{10}, \bibinfo{number}{1} (\bibinfo{year}{2019}),
  \bibinfo{pages}{80--91}.
\newblock


\bibitem[\protect\citeauthoryear{Wood, Robinson, Costa, and Beltran}{Wood
  et~al\mbox{.}}{2021}]%
        {wood2021accuracy}
\bibfield{author}{\bibinfo{person}{Sarah~A Wood}, \bibinfo{person}{Patrick~W
  Robinson}, \bibinfo{person}{Daniel~P Costa}, {and} \bibinfo{person}{Roxanne~S
  Beltran}.} \bibinfo{year}{2021}\natexlab{}.
\newblock \showarticletitle{Accuracy and precision of citizen scientist animal
  counts from drone imagery}.
\newblock \bibinfo{journal}{\emph{PloS one}} \bibinfo{volume}{16},
  \bibinfo{number}{2} (\bibinfo{year}{2021}), \bibinfo{pages}{e0244040}.
\newblock


\end{thebibliography}

\end{document}